\PassOptionsToPackage{table}{xcolor}

\documentclass[table]{article}

\usepackage{arxiv}

\usepackage[utf8]{inputenc} 
\usepackage[T1]{fontenc}    
\usepackage{hyperref}       
\usepackage{url}            
\usepackage{booktabs}       
\usepackage{amsfonts}       
\usepackage{nicefrac}       
\usepackage{microtype}      
\usepackage{lipsum}
\usepackage{graphicx}
\usepackage{tikz}
\usepackage{multicol}
\usepackage[utf8]{inputenc}
\usepackage{caption}
\usepackage{amsmath}
\usepackage{subcaption}
\usepackage{bigstrut}
\usepackage{adjustbox}
\usepackage{float}
\usepackage{array}
\usepackage{multirow}
\usepackage{colortbl}
\usepackage{authblk}

\newcolumntype{P}[1]{>{\centering\arraybackslash}p{#1}} 
\newcolumntype{C}[1]{>{\centering\arraybackslash}m{#1}}

\definecolor{mygray}{rgb}{.906, .902, .902}
\DeclareUnicodeCharacter{2061}{\leftexp{}}
\graphicspath{ {./images/} }
\title{DeepCERES: A Deep learning method for cerebellar lobule segmentation using ultra-high resolution multimodal MRI}

\author[1,*]{Sergio Morell-Ortega }
\author[1]{Marina Ruiz-Perez}
\author[2]{Marien Gadea}
\author[3]{Roberto Vivo-Hernando}
\author[4]{Gregorio Rubio}
\author[5]{Fernando Aparici}
\author[6]{Maria de la Iglesia-Vaya}
\author[7]{Gwenaelle Catheline}
\author[8]{Pierrick Coupé}
\author[1]{José V. Manjón}

\affil[1]{Instituto de Aplicaciones de las Tecnologías de la Información y de las Comunicaciones Avanzadas (ITACA), Universitat Politècnica de València, Camino de Vera s/n, 46022, Valencia, Spain}
\affil[2]{Department of Psychobiology, Faculty of Psychology, Universitat de Valencia, Valencia, Spain}
\affil[3]{Instituto de Automática e Informática Industrial, Universitat Politècnica de València, Camino de Vera s/n, 46022, Valencia, Spain}
\affil[4]{Departamento de matemática aplicada, Universitat Politècnica de València, Camino de Vera s/n, 46022 Valencia, Spain}
\affil[5]{Área de Imagen Médica. Hospital Universitario y Politécnico La Fe. Valencia, Spain}
\affil[6]{Unidad Mixta de Imagen Biomédica FISABIO-CIPF. Fundación para el Fomento de la Investigación Sanitario y Biomédica de la Comunidad Valenciana - Valencia, Spain}
\affil[7]{Univ. Bordeaux, CNRS, UMR 5287, Institut de Neurosciences Cognitives et Intégratives d'Aquitaine, Bordeaux, France}
\affil[8]{CNRS, Univ. Bordeaux, Bordeaux INP, LABRI, UMR5800, in2brain, F-33400 Talence, France}
\affil[*]{Corresponding author email: \texttt{sermoor1@teleco.upv.es}}

\date{} 

\begin{document}
\maketitle
\begin{abstract}
This paper introduces a novel multimodal and high-resolution human brain cerebellum lobule segmentation method. Unlike current tools that operate at standard resolution (1 mm\textsuperscript{3}) or using mono-modal data, the proposed method improves cerebellum lobule segmentation through the use of a multimodal and ultra-high resolution (0.125 mm\textsuperscript{3}) training dataset. To develop the method, first, a database of semi-automatically labeled cerebellum lobules was created to train the proposed method with ultra-high resolution T1 and T2 MR images. Then, an ensemble of deep networks has been designed and developed, allowing the proposed method to excel in the complex cerebellum lobule segmentation task, improving precision while being memory efficient. Notably, our approach deviates from the traditional U-Net model by exploring alternative architectures. We have also integrated deep learning with classical machine learning methods incorporating a priori knowledge from multi-atlas segmentation which improved precision and robustness. Finally, a new online pipeline, named DeepCERES, has been developed to make available the proposed method to the scientific community requiring as input only a single T1 MR image at standard resolution.
\end{abstract}


\newpage
\section{Introduction}
The modern age in the anatomical study of the cerebellum started 120 years ago when Santiago Ramón y Cajal published his first paper with Golgi-impregnated material. His pioneering work marked the inception of an extensive effort to unravel the intricate organisation of the central nervous system \cite{Sotelo2008}. Despite representing a small portion of the total volume of the brain, the cerebellum itself contains approximately 50\% of all the neurons in the brain \cite{Laidi2015}, allowing it to play a crucial role in cognitive, emotional, and behavioural functions, as well as its traditional role in movement coordination. Macroscopically, the cerebellum comprises two cerebellar hemispheres connected by a narrow median vermis. It is connected to the posterior aspect of the brainstem by three symmetrical bundles of nerve fibres called the superior, middle, and inferior cerebellar peduncles. The Schmahmann atlas \cite{Schmahmann2000} presents a standardised nomenclature and unifying terminology. This atlas hierarchically differentiates the regions of the cerebellum into groups by its fissures and then into individual folds called lobules, which are identified by Roman numerals from I to X.  

In the last decades, clinical observations and neuroimaging studies have demonstrated the importance of the cerebellum and its communication with the cerebral cortex in humans \cite{Desmond1998,King2019}. In 1998 it was published a first description of the so-called cerebellar cognitive-affective syndrome \cite{Schmahmann1998} , a constellation of cognitive and behavioural impairments such as problems with abstract reasoning and emotional control. This condition, which some now refer to as Schmahmann’s syndrome, helped establish an appreciation of the cerebellum’s role beyond coordinating movement. Nowadays we know its role has also been evidenced in cognitive operations such as learning \cite{Guell2018}, memory \cite{Desmond1998}, language \cite{Marien2014}, and emotional behaviour\cite{Adamaszek2016}. Hence, by quantifying the structure and function of the cerebellum, researchers gain valuable insights into the role it plays in both motor control and cognitive processes.

The field of its study has been and remains very active. Recent research has highlighted that cerebellar lesions' have profound impact on motor and cognitive functions. Lesions in the sensorimotor cerebellar lobules, particularly the anterior lobule, can lead to movement dysmetria, as seen in cerebellar motor syndrome. Conversely, damage to the cognitive-emotional cerebellar lobules in the posterior lobule can result in dysmetria of thought and emotion, giving rise to cerebellar cognitive affective syndrome. This linkage between the cerebellum's roles in motor control and cognition advances our understanding of cognitive mechanisms and holds promise for therapeutic interventions in behavioural neurology and neuropsychiatry, potentially benefiting conditions like schizophrenia or autism \cite{Schmahmann2019}.

In addition to applications in motor impairments such as cerebellar motor syndrome \cite{Schmahmann2019}, quantification of brain volumes has also been employed in other diseases such as spinocerebellar ataxia or amyotrophic lateral sclerosis \cite{Chandrasekaran2022,Bede2021}. Progress has been made in understanding the involvement of the cerebellum in particular mechanisms of cognition and behavioural and neuropsychiatric neurology, with studies on autism  \cite{Laidi2022}, epilepsy \cite{Warren2022}, schizophrenia or bipolar disorder \cite{Laidi2019}. Its implication in dementia has also been studied by examining cerebellar cortical thickness \cite{McKenna2021} or alterations in white and grey matter in frontotemporal dementia \cite{Clouston2022}. 

Therefore, there is no doubt about the importance and impact of the precise quantification of cerebellum function and anatomy. Specifically, the accurate estimation of cerebellum volumetry through magnetic resonance imaging (MRI) has become an important research line in the last years. Although manual segmentation is considered the gold standard \cite{Huo2019}, it has several drawbacks, such as the inter-rater variability \cite{Bogovic2013}, the expertise required or the time needed to complete the task \cite{Carass2018}. These limitations have led to the need to design semi-automatic or automatic techniques for cerebellum segmentation.

One of the first fully automatic cerebellum segmentation methods was the Spatially Unbiased Infra-tentorial Template (SUIT)\cite{Diedrichsen2006}. It was based on the nonlinear registration of a probabilistic cerebellum atlas to the case to be segmented. Similarly, the Multiple Automatically Generated Templates (MAGeT) brain segmentation algorithm \cite{Park2014} relied on creating a library of templates nonlinearly registered the target image. The final segmentation was achieved by merging multiple segmentations based on a majority voting scheme.

With a slightly different perspective, RASCAL \cite{Weier2014} used an approach based on multi-atlas non-local patch-based label fusion method \cite{Coupe2011} using a priori information via majority voting for label fusion and nonlinear registration. Similarly, another multi-atlas-based method called CERES \cite{Romero2017} was proposed using an ultra-fast patch-matching technique called Optimized PatchMatch ALgorithm (OPAL) \cite{Giraud2016}. An incremental version of this method was later proposed, CERES2 \cite{Carass2018} which improved the results of CERES by adding a patch-based neural network method for systematic error correction named PEC (Patch-based Ensemble Corrector) \cite{pec}. 

A review paper on cerebellum segmentation methods was published in 2018 \cite{Carass2018} summarising the results of the cerebellum segmentation challenge of the MICCAI2017 conference, where CERES2 method was found to be the best-performing method in all categories. It is worth noting that competing methods in this challenge also included Deep Learning methods such as LiviaNET \cite{Dolz2018} and two others U-NET-based  \cite{Ronneberger2015} convolutional neural networks.

More recently, new methods for cerebellum segmentation based on deep learning have been proposed. One of them is named ACAPULCO \cite{Huo2019}, an acronym for Anatomical Parcellation using a U-Net with Locally Constrained Optimization. This method uses a two-step deep learning strategy with two 3D convolutional neural networks (CNNs), one to localise the cerebellum and another to segment it. This method was capable of achieving state-of-the-art results yielding similar accuracy that the state-of-the-art method CERES2. 
Finally, the most recent method for cerebellum lobule segmentation is CEREBNET \cite{Faber2022}, which combines the architecture of FastSurferCNN \cite{Henschel2020}, a 2.5D segmentation network based on U-Net, with an extensive data augmentation approach (i.e. realistic nonlinear deformations to increase anatomical variability) eliminating the need of preprocessing steps, such as spatial normalisation or bias field correction.  

All the methods described above have two significant limitations. Firstly, they exclusively operate with mono-modal data, typically T1, owing to its good contrast between cerebellum grey and white matter. However, this can occasionally result in false positives in peri-cerebellar veins and meninges. Secondly, they all operate at standard resolution, typically no greater than 1 mm\textsuperscript{3}, which appears inadequate due to the intricately convoluted nature of the cerebellum and the presence of partial volume artefacts at this resolution.

Therefore, in this paper, we propose a new cerebellum lobule segmentation method that uses multimodal data (T1 and T2-weighted images) to avoid false positives by adding extra information in the classification process, and that works at ultra-high resolution (0.125 mm\textsuperscript{3}, i.e. 0.5x0.5x0.5 mm\textsuperscript{3}) to improve the segmentation accuracy.

\section{Material and methods}
\label{sec:mat_methods}
For the development of the proposed method two MRI datasets were used. The first one was a subset of the Human Connectome Project (HCP) database \cite{VanEssen2012}, including multimodal T1 and T2 images. Specifically, we randomly selected 75 subjects (from the nearly 1200 dataset images). Those images consisted of T1 and T2 MRI volumes with dimensions of 260x311x260 voxels and a resolution of 0.343 mm\textsuperscript{3} (0.7 mm isotropic resolution). These subjects ranged from 22 to 36 years old (41 women and 34 men). This HCP dataset was used to train our proposed method after a semiautomatic labelling process described in the next section. 

The second dataset consisted of 4857 T1 weighted MRI subjects from a recent lifespan paper \cite{Coupe2017} from various publicly available datasets. This dataset was used for data augmentation to increase the training data variability by including individuals from early infancy to old age. These databases are the following: 

\begin{itemize}
    \item NDAR (N=493): The Database for Autism Research (NDAR) is a national database funded by NIH \href{https://ndar.nih.gov)}{dataset}. This database included samples of different cohorts acquired on 1.5T MRI and 3T scanners. 
\item ABIDE (N=905): The images from the Autism Brain Imaging Data Exchange (ABIDE) \href{http://fcon_1000.projects.nitrc.org/indi/abide/}{dataset} were acquired at 20 different sites on 3T scanner. 
\item ICBM (N=294): The images from the International Consortium for Brain Mapping (ICBM) \href{http://www.loni.usc.edu/ICBM/)}{dataset} were obtained through the LONI website. 
\item OASIS (N=393): The subject images come from the Open Access Series of Imaging Studies (OASIS) \href{http://www.oasis-brains.org}{dataset}. 
\item IXI (N=549): The images from the Information eXtraction from Images (IXI) \href{http://brain-development.org/ixi-dataset}{dataset} consist of normal subjects from 1.5 and 3T scanners. 
\item ADNI (ADNI1 and ADNI2) (N=1649): The images from the Alzheimer’s Disease Neuroimaging Initiative (ADNI) \href{http://adni.loni.usc.edu}{dataset} consisted of subjects from the 1.5T and 3 T MR in collection 1 and 2. These images were acquired at different sites across the United States and Canada. 
\item AIBL (N=338) The Australian Imaging, Biomarkers and Lifestyle (AIBL) \href{http://www.aibl.csiro.au/}{dataset}. The imaging protocol was defined to follow ADNI’s guideline on the 3T scanner and a MPRAGE sequence on the 1.5T scanner. 
\item C-MIND (N=236): All the images of the \href{https://research.cchmc.org/c-mind}{dataset} were acquired at the same site on a 3T scanner. The MRI are 3D T1-weighted MPRAGE high-resolution anatomical scan of the entire brain with a spatial resolution of 1 mm\textsuperscript{3}. 
\end{itemize}
Table \ref{tab:QC3000_description} summarizes the whole dataset in terms of the number of samples and the demographics of each database. 

\begin{table}[H]
  \centering
    \begin{adjustbox}{width=\columnwidth,center}
       \begin{tabular}{c|c|c|cc|ccccccc}
    \hline
    \multirow{2}[2]{*}{\textbf{Data Base}} & \multirow{2}[2]{*}{\textbf{Gender}} & \multirow{2}[2]{*}{\textbf{N}} & \multicolumn{2}{c|}{\multirow{2}[2]{*}{\textbf{Age }}} & \multicolumn{7}{c}{\textbf{Diagnostic}} \bigstrut[t]\\
               &            &            & \multicolumn{2}{c|}{}   & \textbf{CN} & \textbf{ASD} & \textbf{AD} & \textbf{EMCI} & \textbf{LMCI} & \textbf{MCI} & \textbf{SMC} \bigstrut[b]\\
    \hline
    \multirow{2}[2]{*}{\textbf{HCP}} & Male       & 34         & \multicolumn{2}{c|}{26,59 ± 3,28} & 34         &            &            &            &            &            &  \bigstrut[t]\\
               & Female     & 41         & \multicolumn{2}{c|}{27,93 ± 3,66} & 41         &            &            &            &            &            &  \bigstrut[b]\\
    \hline
    \multirow{2}[2]{*}{\textbf{ABIDE}} & Male       & 773        & \multicolumn{2}{c|}{17,95 ± 8,39} & 408        & 365        &            &            &            &            &  \bigstrut[t]\\
               & Female     & 132        & \multicolumn{2}{c|}{16,13 ± 7,67} & 84         & 48         &            &            &            &            &  \bigstrut[b]\\
    \hline
    \multirow{2}[2]{*}{\textbf{ADNI}} & Male       & 907        & \multicolumn{2}{c|}{74 ± 7} & 201        &            & 181        & 144        & 91         & 246        & 44 \bigstrut[t]\\
               & Female     & 742        & \multicolumn{2}{c|}{72 ± 7} & 203        &            & 151        & 113        & 77         & 137        & 61 \bigstrut[b]\\
    \hline
    \multirow{2}[2]{*}{\textbf{AIBL}} & Male       & 160        & \multicolumn{2}{c|}{72,4 ± 6,96} & 112        &            & 18         &            &            & 30         &  \bigstrut[t]\\
               & Female     & 178        & \multicolumn{2}{c|}{74,19 ± 7,74} & 120        &            & 29         &            &            & 29         &  \bigstrut[b]\\
    \hline
    \multirow{2}[2]{*}{\textbf{ICBM}} & Male       & 152        & \multicolumn{2}{c|}{32,2 ± 11,91} & 152        &            &            &            &            &            &  \bigstrut[t]\\
               & Female     & 142        & \multicolumn{2}{c|}{35,4 ± 16,4} & 142        &            &            &            &            &            &  \bigstrut[b]\\
    \hline
    \multirow{2}[2]{*}{\textbf{IRC0}} & Male       & 86         & \multicolumn{2}{c|}{9,68 ± 4,4} & 86         &            &            &            &            &            &  \bigstrut[t]\\
               & Female     & 108        & \multicolumn{2}{c|}{7,81 ± 4,63} & 108        &            &            &            &            &            &  \bigstrut[b]\\
    \hline
    \multirow{2}[2]{*}{\textbf{IXI}} & Male       & 242        & \multicolumn{2}{c|}{46,97 ± 16,56} & 242        &            &            &            &            &            &  \bigstrut[t]\\
               & Female     & 307        & \multicolumn{2}{c|}{50,18 ± 16,31} & 307        &            &            &            &            &            &  \bigstrut[b]\\
    \hline
    \multirow{2}[2]{*}{\textbf{NDAR}} & Male       & 303        & \multicolumn{2}{c|}{13,29 ± 6,18} & 208        & 95         &            &            &            &            &  \bigstrut[t]\\
               & Female     & 190        & \multicolumn{2}{c|}{12,21 ± 5,1} & 174        & 16         &            &            &            &            &  \bigstrut[b]\\
    \hline
    \multirow{2}[2]{*}{\textbf{OASIS}} & Male       & 150        & \multicolumn{2}{c|}{63,91 ± 11,76} & 111        &            & 16         &            &            & 23         &  \bigstrut[t]\\
               & Female     & 243        & \multicolumn{2}{c|}{67,56 ± 13,2} & 187        &            & 29         &            &            & 27         &  \bigstrut[b]\\
    \hline
    \multirow{2}[2]{*}{\textbf{UCLA}} & Male       & 21         & \multicolumn{2}{c|}{7,75 ± 2,42} & 21         &            &            &            &            &            &  \bigstrut[t]\\
               & Female     & 21         & \multicolumn{2}{c|}{7,4 ± 3,04} & 21         &            &            &            &            &            &  \bigstrut[b]\\
    \hline
    \end{tabular}%
    \end{adjustbox}
    \vspace{0.2 cm}
  \caption{Databases description and demographics. (CN: Cognitive normal, ASD: Autism spectrum disorder, AD: Alzheimer's disease, EMCI: early mild cognitive impairment, LMCI: Late mild cognitive impairment, MCI: mild cognitive impairment, SMC: subject memory complaints)}
    \label{tab:QC3000_description}%
\end{table}%

\subsection{Ultra-high resolution cerebellum lobule labelling}

To create our own library with the corresponding ground truth segmentations, we used the CERES method \cite{Romero2017}, considering it one of the best automatic methods based on the comparison of \cite{Carass2018}. It is worth noting that the CERES method's accuracy was reported to be close to the intra-ratter accuracy. 

To generate the HCP data segmentations, the images underwent the specific preprocessing steps of CERES, which are part of the preprocessing in the proposed method, as we will describe later. This preprocessing aims to locate the images in a standardized geometric and intensity space. The preprocessing steps are as follows. First, noise reduction in the native space is performed using the Spatially Adaptive Non-Local Means Filter method \cite{Manjon2010}, followed by an inhomogeneity correction using the N4 method \cite{Tustison2010}. Then, an affine registration to the MNI152 space at a 0.125 mm\textsuperscript{3} resolution is carried out using the ANTS tool \cite{Avants2009} to put the volumes into a common coordinate space. Next, a second inhomogeneity correction was applied only to the brain using the MNI152 intracranial mask using the N4 method. Lastly, a final step involves cropping the cerebellum area to reduce computational requirements. It is worth noting that this process was applied to both T1 and T2 images of the dataset (although only the T1 were used within the CERES method). 

After applying all of these preprocessing steps, the resulting volume had dimensions of 252x156x162 voxels with a resolution of 0.125 mm\textsuperscript{3}. However, CERES method operates at a resolution of 1 mm\textsuperscript{3}. To be able to apply CERES method we employed a stride volume decomposition technique \cite{Manjon2020}. This approach is based on the decomposition into 8 volumes at a 1 mm\textsuperscript{3} resolution of a single 0.125 mm\textsuperscript{3} volume by sampling the 3D volume eight times with a step size of 2 voxels in each dimension with eight different offsets. Subsequently, these 8 volumes were segmented, and finally, the resulting segmentations were mapped back to the original 0.125 mm\textsuperscript{3} space by reversing the stride operation.

\begin{figure}[H]
    \centering
    \includegraphics[width=\textwidth]{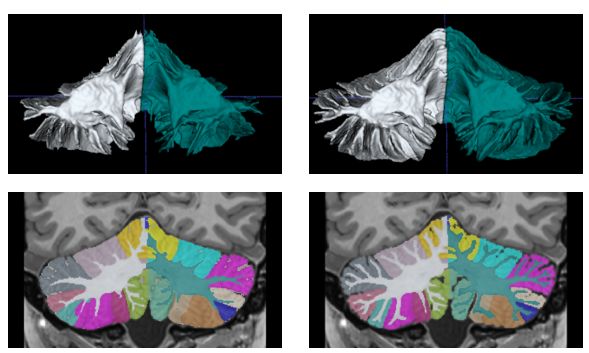}
    \caption{Left: 3D reconstruction of WM label and a coronal view of segmentation before semiautomatic correction. Right: 3D reconstruction of WM label and a coronal segmentation view after semiautomatic correction. The WM “arbor vitae” is better defined in the corrected version.}
    \label{fig:mejora_manual}
\end{figure}

The resulting segmentations had an overall good quality but suffered from three types of errors. First, CSF voxels within neighbor lobules were misclassified as GM since, at 1 mm\textsuperscript{3} resolution, those voxels are affected by partial volume effects and typically not assigned to CSF tissue. Second, similarly, cerebellar white matter (also known as the "tree of life" or "arbor vitae" by anatomists due to its branching structure) was underestimated close to the cerebellar cortex again due to partial volume problems. Finally, some lobule parts were misclassified due to other reasons. The first two problems were partially solved using intensity information of both T1 and T2 images, adding voxels with intensities coherent with the corresponding tissue in each case (CSF or WM). The remaining labelling errors were manually corrected by an expert using ITK-SNAP software \cite{Yushkevich2006}. We also manually removed the cerebellar peduncles from the WM label as we do not consider them as part of the cerebellum. An example of the semiautomatic correction described is shown in Figure \ref{fig:mejora_manual}.
\vspace{0.5 cm}

As a result of the labelling process, a training dataset of 75 T1 and T2 images at 0.125 mm\textsuperscript{3} MNI space with their corresponding labels was generated. To double the size of the training library, the left-right mirrored versions were also included, resulting in a final library of 150 cases. This library was divided into 130 cases for training, 10 cases for validation and 10 cases for test.

\subsection{Segmentation method}

The challenges of segmenting 27 cerebellar structures with ultra-high resolution multimodal volumes is the computational complexity and the memory consumption due to 3D convolutional neural networks. Thus, the right choice of network architecture plays a crucial role in the method's implementation. The U-net-based \cite{Ronneberger2015} architecture has been the most used topology in medical image analysis tasks. Previous methods for cerebellum segmentation have used 2.5D or 3D U-net versions, being the last one the preferred option due to its better context analysis, which usually improves classification accuracy. Unfortunately, at 0.125 mm\textsuperscript{3} resolution, a 3D U-net is not a feasible option due to the memory limitations of currently available GPUs. 

\begin{figure}[H]
    \centering
    \includegraphics[width=\textwidth]{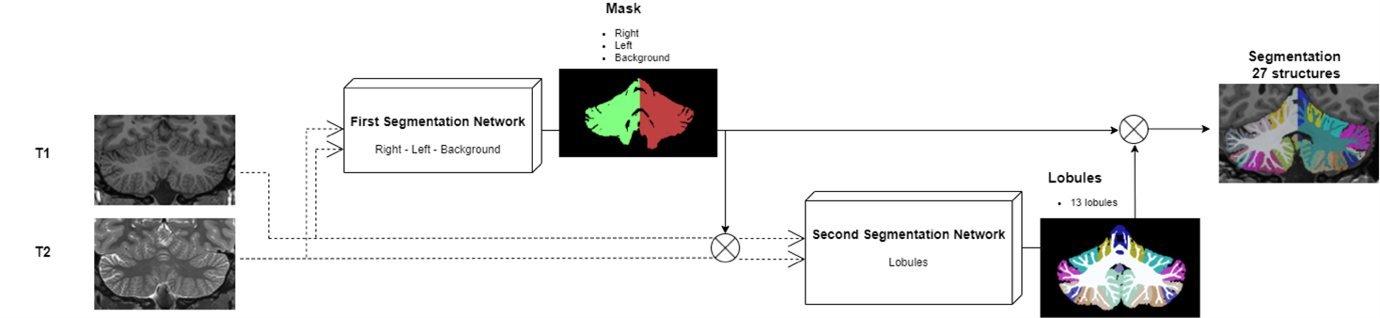}
    \caption{Scheme of the proposed segmentation method. The first network segments left-right cerebellum and background and the second network segments the lobules using the T1 and T2 images and the output of the first network.}
    \label{fig:cascada}
\end{figure}

To deal with this limitation while maintaining the 3D nature of the proposed method, a two-stage cascade architecture is proposed, taking advantage of the pseudo-symmetry of the human cerebellum. Firstly, the left and right hemispheres are segmented using a first network, allowing the generation of a binary mask that classifies voxels into the left-right cerebellum and the background. Subsequently, a binary mask cerebellum/background from the output of this first network is multiplied to the T1 and T2 volumes to generate the input of a second network, thereby facilitating intra-cerebellar attention of the 14 classes (12 lobules, WM and background) segmentation network (regardless of whether they are left or right). Finally, to generate the 27 labels segmentation, outputs from both networks are combined to identify each hemisphere's left and right structures. The scheme of this process is summarized in Figure \ref{fig:cascada}. 
\vspace{0.5 cm}

We initially used for both networks a classical U-net 3D architecture as reference model. However, even with the cascade approach, the maximum number of filters for the 3D U-net was limited to 16 to fit into memory. 

\begin{figure}[H]
  \centering
  \resizebox{\textwidth}{!}{
\input{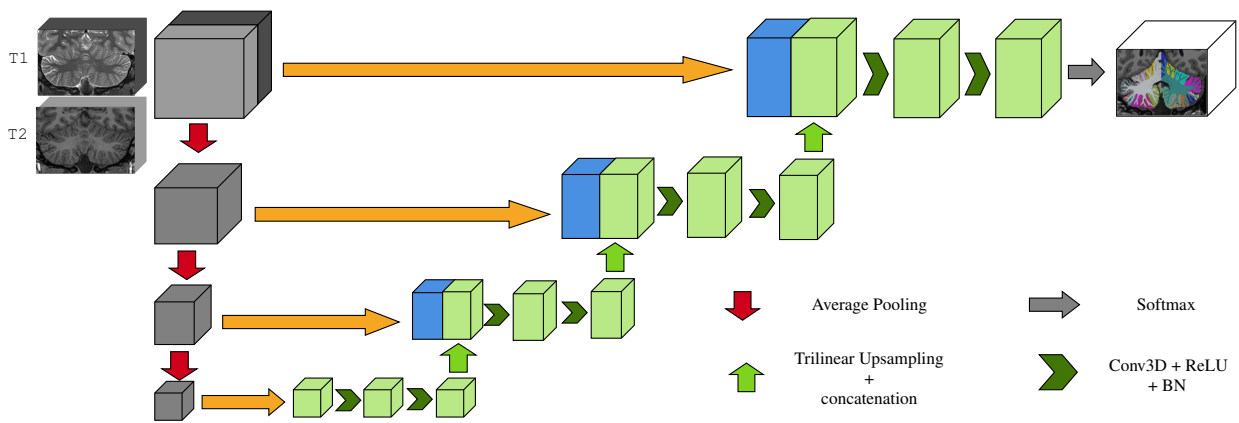}}
  \caption{Proposed DPN architecture.}
  \label{fig:DPN}
\end{figure}

To mitigate complexity, we explored the design of a more streamlined novel network characterized by fewer parameters. This approach aims to facilitate effective learning of segmentations while mitigating the risk of succumbing to overfitting or underfitting. Therefore, we experimented with a self-designed architecture based on a coarse-to-fine approach called DPN (Deep Pyramidal Network), which enables a lighter and less memory demanding model. Unlike U-net architecture, where an encoder-decoder strategy is used, the DPN network uses an 8 times subsampled version of the original volume as initial input (see Figure \ref{fig:DPN}). It concatenates the generated features obtained through three blocks (convolution + ReLU + Bach Normalization) with higher-resolution features obtained from the upper-resolution level to refine frequency details. This process is repeated until the original resolution is reached, and a final softmax layer is used to generate output class probabilities. At the end of each resolution block, a dropout layer was added for the resolution 1/8, 1/4 and 1/2 to minimize overfitting problems (rate 0.25). This architecture reduces the number of parameters to almost one-fifth compared to a U-net model. A scheme of the proposed DPN architecture is shown in Figure \ref{fig:DPN}.        
\vspace{0.5 cm}

In the training process, a loss function composed of binary cross-entropy and Dice loss is minimized. To prevent gradient vanishing problems when the values of the loss function are close to 0, the logarithm operation is applied to the loss function to scale the gradients. 
\begin{equation}
    L = \log\left(\frac{1 - \sum_{i=1}^{C} \text{Dice}(y_{c},\hat{y}_{c}) + \text{BCE}(y,\hat{y})}{2}\right)
\end{equation}

Where $C$ represents the classes (labels), Dice stands for the Dice index, and BCE is the Binary Cross-Entropy.

\subsection{Atlas guided segmentation}

On the other hand, we realized that in our lobule segmentation problem, different lobules have similar intensity patterns but different locations with sometimes no clear limits between neighbor lobules. Atlas-based methods \cite{Carass2018} benefit from the location information by locally searching for similar patches (for example CERES method) after a registration process. Therefore, we hypothesized that including a priori information (i.e., a subject-specific atlas) could assist in the segmentation process, reducing labelling errors and making the method more robust. Using this a priori information could reduce the sensitivity of the models to noise, artefacts or variability of the input data, avoiding catastrophic errors and non-plausible topologies. 

A multi-atlas strategy is used to create this atlas, as it has proven more robust than the single-atlas. In this approach, a nonlinear registration transform is estimated for each anatomical volume (T1) in the training library to the volume to be segmented. Subsequently, these non-linear transforms are applied to the volumes and their corresponding segmentations, resulting in a subject-specific library of segmentations. Finally, these segmentations are merged voxel-wise according to a rule that assigns a value (label) to each voxel in the volume. 

Nonetheless, the primary limitation of this methodology, as previously identified in the article by \cite{Park2014}, is the time cost associated with non-linear registrations and label fusion. Fortunately, deep learning-based methods, such as VoxelMorph, have drastically reduced the time cost of non-linear registration from minutes to less than one second \cite{Balakrishnan2018}. In the proposed method, we used a similar deep non-linear registration network than\cite{Coupe2020}. This replaces the costly registration process by evaluating a convolutional neural network previously trained on the training cerebellum dataset, substituting a computationally expensive minimization process with a memory-based one, with a time cost of 0.7 seconds per cerebellar volume.

To perform the label fusion, an intensity-based weighted majority voting is used where voxels of the library with intensities similar to the voxel to be classified have a higher weight than those being more dissimilar. Each weight is calculated as follows:

\begin{equation}
    w_p = \frac{1}{1 + d\cdot|\lvert I_p - L_{p,j}| \rvert}
\end{equation}

Where $d$ is a scaling coefficient, $I_p$ is the intensity of a voxel at a position $p$, and $L_p$,$j$ is the intensity of a voxel at a position $p$ of an image $j$ from library $L$. We set $d$ as 0.5 experimentally. 

To deal with the temporal limitations of multi-atlas approaches, the fusion algorithm was implemented in C language and parallelized using the OpenMP multiprocessing library \cite{openMP}. Another aspect that impacts the temporal cost of the atlas generation process is the number of templates used to generate it (the bigger the library, the more time will be spent on the atlas creation). Therefore, given the availability of 130 T1/T2 images with their corresponding segmentations in the library, we select only the N cases that are more similar (based on the L1 norm of image intensities) to perform the non-linear registration and the later label fusion. 

To determine the optimal number of elements in the library, we used the 130 segmentations, from which we extracted a sample to calculate the specific atlas (leave-one-out). Empirically, it is observed that a plateau is reached in terms of the Dice between the atlas and the original segmentation (0.82) when the size of the library is N=20 samples more similar to the sample under analysis. 

\subsection{Extensive data augmentation}

Since our models were trained in an age-limited training dataset, we decided to augment the training data to include more diverse anatomy patterns. To extent the training dataset in a realistic manner without having more manually labelled cases we used an approach based on non-linear registration. We used the lifespan dataset (N=4857) cases to obtain a diverse shape population from samples from childhood to old age. 

Specifically, for each case of the lifespan dataset we look for the most similar case (T1) in our semiautomatically labelled training dataset and this case is non-linear registered to the reference lifespan case using Greedy algorithm \cite{Battle2016}. Once the new 4857 new cases are generated (T1,T2 and Atlas) a fine tunning process is performed using as training data, both, the original training set (130 samples) and the extended dataset (4857 samples) with a 50\% of probability each to balance both datasets.

\section{Experiments and results}

In this section, we show the results of several experiments performed to determine the best configuration of the proposed method. All the experiments have been performed using the TensorFlow 1.15 framework, Keras 2.2.4, and a batch size of 1 on a GPU V100 with 32 GB of memory. For training Adam optimizer was used during the first 2000 epochs and then Adamax \cite{King2019} is used for an additional 1000 epochs. In the case of finetuning process, Adamax was directly used during 1000 epochs. As image inputs, we used the T1 and T2 volumes normalized to have mean 0 and variance 1 (subtracting the mean and dividing by the standard deviation).

To further increase input data variability during the first part of the training (HCP data only), we have used random data augmentation through the Torchio library \cite{Perez-Garcia2021}, enabling the online generation of typical transformations within the medical image domain. Intensity transformations such as random anisotropy, bias field, blur, ghosting and gamma have been applied alongside geometric transformations, including random affine and elastic deformation.

\subsection{U-Net vs DPN}
Both network architectures were trained independently to evaluate the choice of the optimal architecture (U-net vs DPN). Two configurations were considered within a cascade framework (2 DPN and 2 U-NET) for right-left background and lobule networks. For both networks in the cascade, the number of filters of DPN was set to 32 for all resolutions, while for U-net, the initial number of filters was set to 16, doubling this number at each down-sampling step. As shown in Table \ref{tab:dpn_vs_unet}, both models present a similar performance (the U-net based option is slightly superior, although with a smaller standard deviation, while being 5 times bigger). 

\begin{table}[htbp]
  \centering
    \begin{adjustbox}{width=\columnwidth,center}
    
    \begin{tabular}{C{0.22\linewidth}| C{0.22\linewidth}C{0.22\linewidth}C{0.22\linewidth}C{0.22\linewidth}}
    \hline
    \textbf{Model architecture} & \textbf{Parameters} & \textbf{Mean lobule Dice} & \textbf{Mean whole cerebellum Dice} & \textbf{Mean computational time (sec)} \bigstrut[b]\\
    \hline
    2 DPN      & 696177     & 0.9254 (0.036) & 0.9890 (0.0022) & 6.3029 (1.646) \bigstrut[t]\\
    2 U-net    & 3433473    & 0.9286 (0.042) & 0.9911 (0.0028) & 7.5103 (1.611) \bigstrut[b]\\
    \hline
    \end{tabular}%
    \end{adjustbox}
    \vspace{0.1 cm}
    \caption{DPN and U-net architecture comparison. Mean Values (Standard Deviation).}
  \label{tab:dpn_vs_unet}%
\end{table}%

\subsection{ Use of a subject-specific atlas as an additional channel}

Once both architectures were validated, we decided to train both architectures with an additional input channel in the form of a subject-specific atlas to validate the suitability of introducing a priori information into the networks. Table \ref{tab:atlas_vs_noatlas} shows the results of the U-net and DPN networks.

\begin{table}[htbp]
  \centering
    \begin{adjustbox}{width=\columnwidth,center}
    \begin{tabular}{C{0.33\linewidth}|C{0.33\linewidth}C{0.33\linewidth}}
    \hline
    \textbf{Model architecture} & \textbf{Mean lobule Dice} & \textbf{Mean whole cerebellum Dice} \bigstrut\\
    \hline
    2 DPN without atlas & 0.9254 (0.036) & 0.9890 (0.0028) \bigstrut[t]\\
    2 DPN with atlas & 0.9308 (0.042) & 0.9905 (0.0015) \\
    2 U-net without atlas & 0.9286 (0.042) & \textbf{0.9911 (0.0022)} \\
    2 U-net with atlas & \textbf{0.9322 (0.039)} & 0.9898 (0.0020) \bigstrut[b]\\
    \hline
    \end{tabular}%
    \end{adjustbox}
    \vspace{0.1 cm}
    \caption{Atlas vs non-atlas DPN and U-net architecture comparison. Best results in bold. Mean Values (Standard Deviation)}
  \label{tab:atlas_vs_noatlas}%
\end{table}%

As expected, both architectures benefited from the inclusion of a priori information. The DPN-based model obtained the better results for whole cerebellum, while the U-net-based model got the best lobule segmentation Dice. Based on these results, we further investigated the results in Section \ref{sec:ensemble}. 

\subsection{Multimodality and ultra-high resolution impact} 

To validate the original hypothesis on the impact of using ultra-high resolution images (0.125 mm\textsuperscript{3}) and multimodality, both architectures were re-trained following the same strategy (3 input channels and cascade approach), but changing the input data to the networks using only monomodal and/or standard resolution (SR) data. Results can be analyzed in Table \ref{tab:modality_SR}.

\begin{table}[H]
  \centering
    \begin{adjustbox}{width=\columnwidth,center}
    \begin{tabular}
{C{0.25\linewidth}|C{0.25\linewidth}C{0.25\linewidth}C{0.25\linewidth}}
    \hline
    \textbf{Resolution / Multimodality} & \textbf{Model} & \textbf{Structure Dice} & \textbf{Whole Cerebellum Dice} \bigstrut\\
    \hline
     \textbf{SR, T1} & DPN        & 0.9117     & 0.9889 \bigstrut[t]\\
    \textbf{SR, T1} & U-net      & 0.9079     & 0.9878 \\
     \textbf{SR, T1+T2} & DPN        & 0.9138     & 0.9892 \\
     \textbf{SR, T1+T2} & U-net      & 0.9088     & 0.9884 \\
    \textbf{HR, T1} & DPN        & 0.9239     & 0.9896 \\
    \textbf{HR, T1} & U-net      & 0.9276     & 0.9902 \\
    \textbf{HR, T1 + T2} & DPN        & 0.9308     & \textbf{0.9905} \\
    \textbf{HR, T1 + T2} & U-net      & \textbf{0.9322} & 0.9898 \bigstrut[b]\\
    \hline
    \end{tabular}%
    \end{adjustbox}
    \vspace{0.1 cm}
    \caption{Impact of modality and resolution on segmentation results. (HR) High Resolution and (SR) Standard Resolution. Best results in bold.}
  \label{tab:modality_SR}%
\end{table}

On the one hand, the increased resolution has a significant impact on the Dice index at the lobule level (0. 9322 vs 0.9088 in U-net and 0.9308 vs 0.9138 in DPN) and somewhat less as in the whole cerebellum (0.9904 vs 0.9884 in U-net and 0.9907 vs 0.9892 in DPN). This can be seen in the trend of the Dice index with increasing resolution and the introduction of the channel in T2. It is worth remembering the intricate structure of the cerebellum, where there is a high degree of tissue packing and a highly textured surface, the folia, with submillimeter cortical thicknesses. This causes partial volume problems in standard resolution images (1 mm\textsuperscript{3}) where different tissues provide information on the same voxel, preventing their independent classification. The partial volume problems are reduced at ultra-high resolution, allowing the model to extract more information about the tissues and their textures as they are better differentiated. In addition, better performance is observed for the U-net-based structure for HR (but not for SR), under the hypothesis that having a higher capacity (a greater number of parameters) than the DPN-based structure can extract more information from the input data. 

On the other hand, the introduction of T2 benefits to a lesser extent than the increase in resolution (0.9321 vs. 0.9276 in U-net and 0. 9308 vs. 0.9239 in DPN). Although the gain is smaller, it reduces some false positives and negatives by providing information from another modality for the same anatomical structure.

\subsection{Robustness study}

Traditionally, when it comes to ensuring the robustness of the network against anomalous or low-quality data, it is common to employ an extensive process of data augmentation techniques involving transformations applied to the images. In the same line, we believe that the presence of an atlas generated from a fixed library promotes the consistency of the input data, thus increasing the robustness of the results.

To analyze this aspect on the proposed architectures, we studied the impact of the Dice coefficient on models trained with and without an atlas as additional input, observing a lesser loss in performance (almost half) when prior information is available (see Table \ref{tab:robustness_t1}).

\begin{table}[H]
  \centering
    \begin{adjustbox}{width=\columnwidth,center}
    \begin{tabular}{C{0.33\linewidth}|C{0.33\linewidth}C{0.33\linewidth}}
    \hline
    \multicolumn{1}{c|}{\textbf{Name}} & \textbf{Mean Dice} & \textbf{ $\Delta$ Mean Dice} \bigstrut\\
    \hline
      DPN + ATLAS + bad T2 & 0.82552    & \multirow{2}[2]{*}{10.8368} \bigstrut[t]\\
      DPN + ATLAS + original T2 & 0.93388    &  \bigstrut[b]\\
    \hline
      U-net + ATLAS + bad T2 & 0.878601   & \multirow{2}[2]{*}{5.5946} \bigstrut[t]\\
      U-net + ATLAS + original T2 & 0.934547   &  \bigstrut[b]\\
    \hline
    DPN + bad T2 & 0.706626   & \multirow{2}[2]{*}{21.8804} \bigstrut[t]\\
    DPN + original T2 & 0.92543    &  \bigstrut[b]\\
    \hline
    U-net + bad T2 & 0.834663   & \multirow{2}[2]{*}{9.3977} \bigstrut[t]\\
    U-net + original T2 & 0.92864    &  \bigstrut[b]\\
    \hline
    \end{tabular}%
    \end{adjustbox}
    \vspace{0.1 cm}
    \caption{Results of the different robustness settings. Loss of performance when having a bad T2.}
  \label{tab:robustness_t1}%
\end{table}

On the one hand, we simulated the presence of a poor-quality T2 image since it's often easier to have a T1 image than a T2 image. To mimic a bad T2, we replaced the T2 image with the T1 image over the test. Results in Table \ref{tab:robustness_t1} clearly show that networks with an atlas have approximately half the Dice loss compared to models without the atlas as an additional channel. Besides, these results also reveal that U-net is more robust than DPN, with a smaller loss in terms of Dice that can be explained by the higher number of parameters.


On the other hand, the models are subjected to random transformations within the medical imaging domain using the Torchio library \cite{Perez-Garcia2021} for testing its robustness against them. It should be noted that all the models have undergone transformations of this type in the training process as a data augmentation technique. Table \ref{tab:robustness_torchio} presents the robustness results of the Dice coefficient when perturbing input data with random augmentations at test time. As can be noticed, the models with atlas are more robust than their non-atlas versions.

\begin{table}[H]
  \centering
    \begin{adjustbox}{width=\columnwidth,center}
    \begin{tabular}{C{0.15\linewidth}|C{0.1\linewidth}C{0.1\linewidth}C{0.1\linewidth}C{0.1\linewidth}C{0.1\linewidth}C{0.1\linewidth}|C{0.15\linewidth}}
    \hline
    \multicolumn{1}{c|}{\multirow{2}[1]{*}{\textbf{Model}}} & \textbf{Random} & \textbf{Random} & \textbf{Random} & \textbf{Random} & \textbf{Random} & \textbf{Random} & \multirow{2}[1]{*}{\textbf{Mean }} \bigstrut[t]\\
    \multicolumn{1}{c|}{} & \textbf{Anisotropy} & \textbf{Bias Field} & \textbf{Blur} & \textbf{Deformation} & \textbf{Gamma} & \textbf{Ghosting} &  \\
    \hline
    2 DPN with atlas & 0.8817     & 0.6835     & 0.8996     & \textbf{0.8887} & 0.9335     & 0.8719     & 0.8346 \\
    2 Unet with atlas  & \textbf{0.8862} & \textbf{0.7114} & \textbf{0.9039} & 0.8881     & \textbf{0.9344} & \textbf{0.8793} & \textbf{0.8408} \\
    2 U-nets   & 0.8786     & 0.545      & 0.8979     & 0.8846     & 0.9282     & 0.8661     & 0.8243 \\
    2 DPN      & 0.8765     & 0.6335     & 0.8949     & 0.8868     & 0.9248     & 0.8652     & 0.8352 \bigstrut[b]\\
    \hline
    \end{tabular}%
    \end{adjustbox}
    \vspace{0.1 cm}
    \caption{Dice coefficient of the proposed method in function of the augmentation type on the test data. Best results in bold.}
  \label{tab:robustness_torchio}%
\end{table}

\subsection{Ensemble of topologies}
\label{sec:ensemble}
As shown in Table \ref{tab:atlas_vs_noatlas}, the U-Net model had better mean lobule Dice while the DPN model had better whole cerebellum Dice. Deeply analyzing the differences among models, we found some complementarity for different lobules, where sometimes one of the networks is better than the other (see Table \ref{tab:dice_ensemble}). Therefore, instead of choosing one model, we combined both by averaging their predictions using a classical bagging approach. Ensemble learning is a widely recognized concept in machine learning and is employed to improve the overall performance of a technique. This improvement is achieved by training multiple models before combining them \cite{Kamnitsas2018}.

Table \ref{tab:dice_ensemble} shows the results of the ensemble of U-Net and DPN models. As can be noticed, the results of the ensemble for each individual lobule are better than any of the original networks.

\begin{table}[htbp]
  \centering
    \begin{adjustbox}{width=\columnwidth,center}
    \begin{tabular}{C{0.25\textwidth}|C{0.25\textwidth}C{0.25\textwidth}C{0.25\textwidth}}
      \hline
    \multicolumn{1}{C{0.25\textwidth}}{\textbf{Label}} & \multicolumn{1}{C{0.25\textwidth}}{\textbf{2 U-nets with atlas }} & \multicolumn{1}{C{0.25\textwidth}}{\textbf{2 DPN with atlas }} & \multicolumn{1}{C{0.25\textwidth}}{\textbf{Ensemble }} \bigstrut[b]\\
    \hline
    \hline
    Whole cerebellum  & 0.9898 (0.0025) & 0.9905 (0.0024) & \textbf{0.9906 (0.0024) }\bigstrut\\
    \hline
    Right Lobule I-II & 0.8333 (0.0614) & 0.8303 (0.0588) & \textbf{0.8426 (0.0484)} \bigstrut[t]\\
    Right Lobule III & 0.9104 (0.0237) & 0.9107 (0.0238) & \textbf{0.9157 (0.0217)} \\
    Right Lobule IV & 0.9213 (0.0108) & 0.9258 (0.0102) & \textbf{0.9309 (0.0083)} \\
    Right Lobule V & 0.9257 (0.0131) & 0.9295 (0.0167) & \textbf{0.9352 (0.0142)} \\
    Right Lobule VI & 0.9536 (0.0083) & 0.9552 (0.0091) &\textbf{ 0.9579 (0.0078)} \\
    Right Crus I & 0.9597 (0.0072) & 0.9608 (0.0059) & \textbf{0.9629 (0.0063)} \\
    Right Crus II & 0.9502 (0.0122) & 0.9502 (0.0104) & \textbf{0.9541 (0.0105)} \\
    Right Lobule VIIB & 0.9277 (0.0240) & 0.9288 (0.0215) & \textbf{0.9362 (0.0206)} \\
    Right Lobule VIIIA & 0.9421 (0.0183) & 0.9428 (0.0185) & \textbf{0.9480 (0.0178)} \\
    Right Lobule VIIIB & 0.9459 (0.0085) & 0.9418 (0.0102) & \textbf{0.9479 (0.0080)} \\
    Right Lobule IX & 0.9459 (0.0217) & 0.9420 (0.0267) & \textbf{0.9481 (0.0212) }\\
    Right Lobule X & 0.9470 (0.0101) & 0.9440 (0.0114) & \textbf{0.9488 (0.0100)} \\
    Right Grey Matter & 0.9597 (0.0031) & 0.9601 (0.0037) & \textbf{0.9622 (0.0031)} \\
    Left Lobules I-II & 0.8245 (0.0561) & 0.8166 (0.0869) & \textbf{0.8259 (0.0717)} \\
    Left Lobule III & 0.9159 (0.0183) & 0.9137 (0.0185) & \textbf{0.9197 (0.0188)} \\
    Left Lobule IV & 0.9201 (0.0127) & 0.9184 (0.0157) & \textbf{0.9236 (0.0140)} \\
    Left Lobule V & 0.9212 (0.0149) & 0.9160 (0.0201) & \textbf{0.9267 (0.0159)} \\
    Left Lobule VI & 0.9533 (0.0079) & 0.9534 (0.0082) & \textbf{0.9577 (0.0065)} \\
    Left Crus I & 0.9586 (0.0062) & 0.9592 (0.0079) & \textbf{0.9614 (0.0070)} \\
    Left Crus II & 0.9544 (0.0051) & 0.9536 (0.0060) & \textbf{0.9573 (0.0050)} \\
    Left Lobule VIIB & 0.9348 (0.0151) & 0.9284 (0.0227) & \textbf{0.9365 (0.0191)} \\
    Left Lobule VIIIA & 0.9389 (0.0171) & 0.9338 (0.0243) & \textbf{0.9403 (0.0200)} \\
    Left Lobule VIIIB & 0.9399 (0.0107) & 0.9377 (0.0119) & \textbf{0.9432 (0.0103)} \\
    Left Lobule IX & 0.9474 (0.0127) & 0.9437 (0.0145) & \textbf{0.9493 (0.0126)} \\
    Left Lobule X & 0.9442 (0.0129) & 0.9432 (0.0122) & \textbf{0.9465 (0.0121)} \\
    Left White Matter & 0.9612 (0.0026) & 0.9614 (0.0027) & \textbf{0.9633 (0.0027)} \bigstrut[b]\\
    \hline
    Average    & 0.9322$^{**}$ (0.0392) & 0.9308$^{**}$ (0.0427) & \textbf{0.9362 (0.0388)} \bigstrut[t]\\
    \end{tabular}%
    \end{adjustbox}
    \vspace{0.1 cm}
    \caption{Dice results for each lobule for the U-net, DPN and Ensemble models. ($^{**}$ significant difference of the individual models compared to the ensemble model with p<0.01, two-sided Wilcoxon test).}
  \label{tab:dice_ensemble}%
\end{table}

\subsection{Fine tuning with realistic data augmentation}

As commented previously, in order to make the proposed method more robust to anatomy variations we performed a fine-tuning of the trained networks using a realistically augmented dataset using non-linear registration using samples from the lifespan dataset. This fine-tuning was applied to both tasks: right-left-background hemisphere segmentation and lobule segmentation, using Adamax during 1000 epochs. To mitigate the risk of forgetting, cases from the original library are intermittently presented throughout the training process with a 50\% probability, alongside an equal 50\% probability of cases from the extended library. Results of the fine-tuning are shown in Table \ref{tab:dice_ensemble_after_qc3000}. The table shows that there are no significantly different values for any of the labels, demonstrating that the fine-tuning process has not altered the performance of the original method.

As can be noticed, the test results after fine-tuning are slightly lower than before fine-tuning. This is probably due to the fact that improving the generalization makes the method less efficient in the age limited test set (maybe due to overfitting to the HCP dataset). In other words, the improved generalization has a price to pay in form of lower results on the specific test dataset. 

\begin{figure}[H]
    \centering
    \includegraphics[width=\textwidth]{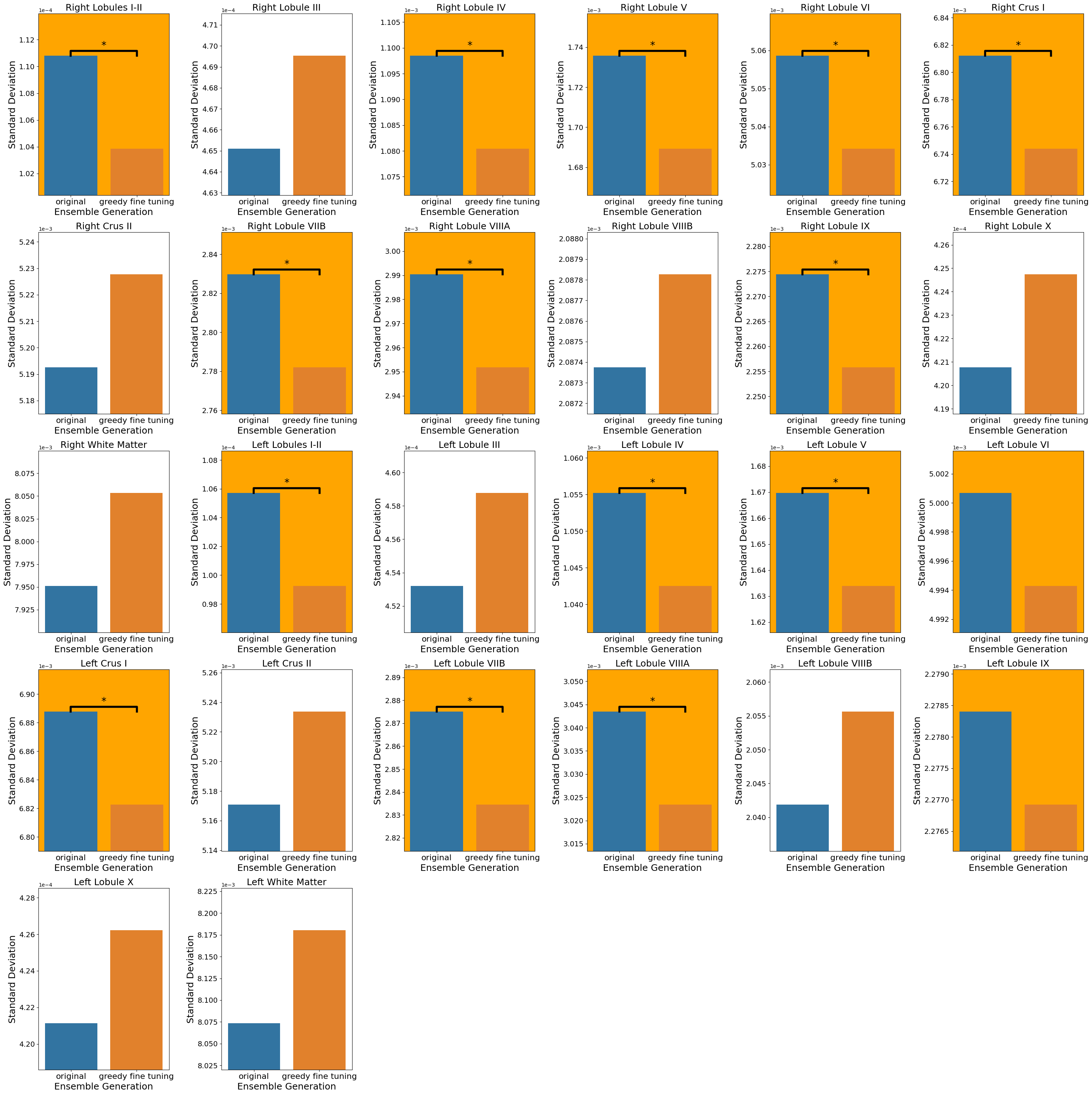}
    \caption{Standard deviation of the volumes measured for each structure. * significant differences in the mean values of standard deviation of both methods when using the Wilcoxon two-sided test with p<0.05.}
    \label{fig:stats}
\end{figure}

To quantitatively measure the generalization gain of the fine-tuning process, we used an indirect approach. Since we have no ground truth labels for the lifespan dataset, we decided to use a population-based validation approach. Specifically, we selected around 3000 healthy cases from the lifespan dataset and segmented them with the “before” and “after” fine-tuning models to generate age-volume curves for each structure. These data were fitted to their corresponding lifespan models with their corresponding bounds as done in \cite{Coupe2017}. Our hypothesis was that the improved generalization will result in more accurate segmentations for each age range and structure thus reducing the age-related variability of the derived volumes models through the obtention of more coherent results. Results of this analysis showed that 16 out of the 26 GM structures had a reduced variability, with statistical significance observed in 15 (p<0.05) Wilcoxon two-sided test. Details are shown in Figure \ref{fig:stats} in Section \ref{sec:Appendix}. 

\begin{table}[H]
  \centering
    \begin{adjustbox}{width=\columnwidth,center}
    \begin{tabular}{C{0.33\linewidth}|C{0.33\linewidth}C{0.33\linewidth}}
      \hline
    \multicolumn{1}{C{0.33\linewidth}}{\textbf{Label}} & \textbf{Ensemble } & \multicolumn{1}{C{0.33\linewidth}}{\textit{\textbf{Ensemble after fine-tuning }}} \bigstrut[b]\\
    \hline
    \hline
    Whole cerebellum  & 0.9906 (0.0024) & 0.9900 (0.0025) \bigstrut\\
    \hline
    Right Lobule I-II & 0.8426 (0.0484)  & 0.8319 (0.0551) \bigstrut[t]\\
    Right Lobule III & 0.9157 (0.0217) & 0.9132 (0.0218) \\
    Right Lobule IV & 0.9309 (0.0083) & 0.9283 (0.0052) \\
    Right Lobule V & 0.9352 (0.0142) & 0.9317 (0.0130) \\
    Right Lobule VI & 0.9579 (0.0078) & 0.9558 (0.0079) \\
    Right Crus I & 0.9629 (0.0063) & 0.9606 (0.0072) \\
    Right Crus II & 0.9541 (0.0105) & 0.9516 (0.0122) \\
    Right Lobule VIIB & 0.9362 (0.0206) & 0.9319 (0.0232) \\
    Right Lobule VIIIA & 0.9480 (0.0178) & 0.9441 (0.0185) \\
    Right Lobule VIIIB & 0.9479 (0.0080) & 0.9453 (0.0085) \\
    Right Lobule IX & 0.9481 (0.0212) & 0.9443 (0.0263) \\
    Right Lobule X & 0.9488 (0.0100) & 0.9452 (0.0083) \\
    Right Grey Matter & 0.9622 (0.0031) & 0.9605 (0.0029) \\
    Left Lobules I-II & 0.8259 (0.0717) & 0.8234 (0.0442) \\
    Left Lobule III & 0.9197 (0.0188) & 0.9159 (0.0165) \\
    Left Lobule IV & 0.9236 (0.0140) & 0.9206 (0.0140) \\
    Left Lobule V & 0.9267 (0.0159) & 0.9240 (0.0136) \\
    Left Lobule VI & 0.9577 (0.0065) & 0.9556 (0.0065) \\
    Left Crus I & 0.9614 (0.0070) & 0.9600 (0.0061) \\
    Left Crus II & 0.9573 (0.0050) & 0.9544 (0.0053) \\
    Left Lobule VIIB & 0.9365 (0.0191) & 0.9319 (0.0207) \\
    Left Lobule VIIIA & 0.9403 (0.0200) & 0.9370 (0.0214) \\
    Left Lobule VIIIB & 0.9432 (0.0103) & 0.9392 (0.0107) \\
    Left Lobule IX & 0.9493 (0.0126) & 0.9459 (0.0136) \\
    Left Lobule X & 0.9465 (0.0121) & 0.9436 (0.0121) \\
    Left White Matter & 0.9633 (0.0027) & 0.9619 (0.0026) \bigstrut[b]\\
    \hline
    Average    & 0.9362 (0.0388) & 0.9330 (0.0386) \bigstrut\\

    \end{tabular}%
    \end{adjustbox}
    \vspace{0.1 cm}
    \caption{Test Dice before and after fine-tuning with the extended dataset.}
  \label{tab:dice_ensemble_after_qc3000}%
\end{table}

\subsection{Comparison with state-of-the-art methods}
Regarding the comparison with other related methods, we cannot make a direct quantitative comparison because either the libraries or the resolution of the input data is different. However, by doing a pseudo-quantitative comparison of the mean structure Dice reported at each publication, it is possible to observe in Table \ref{tab:comparison_SOTA} a clear gain in the accuracy of the proposed method. 

\begin{table}[H]
  \centering
    \begin{adjustbox}{width=\columnwidth,center}
    \begin{tabular}{C{0.25\linewidth}C{0.25\linewidth}C{0.25\linewidth}C{0.25\linewidth}}
    \hline
    \textbf{Acapulco} & \textbf{Cerebnet } & \textbf{Ceres} & \multicolumn{1}{C{0.25\linewidth}}{\textbf{Deep Ceres }} \bigstrut\\
    \hline
    0.77       & 0.87       & 0.7729     & 0.933 \bigstrut[t]\\
    \end{tabular}%
    \end{adjustbox}
    \vspace{0.1 cm}
    \caption{Comparison of the mean Dice coefficient of the cerebellum lobules with related state of the art methods.}
  \label{tab:comparison_SOTA}%
\end{table}

\subsection{DeepCERES pipeline}
\begin{figure}[H]
    \centering
    \includegraphics[width=\textwidth]{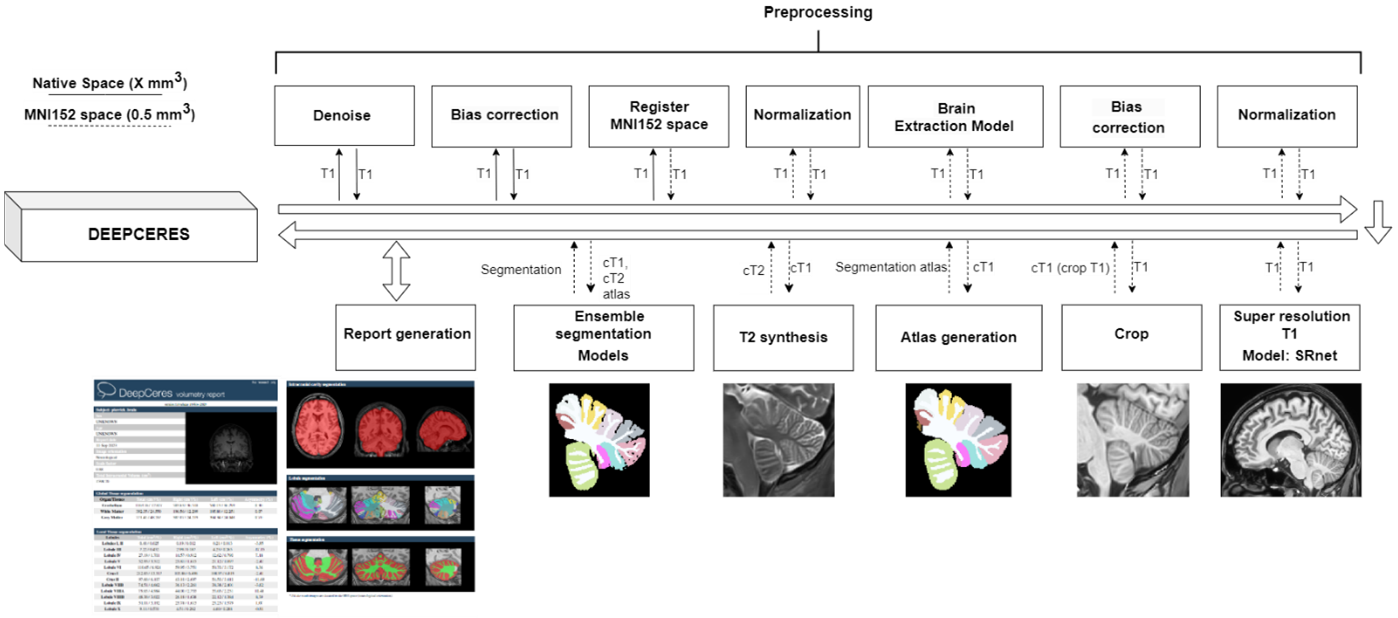}
    \caption{Scheme of the DeepCERES pipeline. }
    \label{fig:pipeline}
\end{figure}

Despite the proposed method good performance with high-resolution multimodal data, we are aware of the limitations of the availability of such data in research and clinical settings. Therefore, we decided to develop a full pipeline able to work on standard resolution T1 MR images. We named the proposed pipeline DeepCERES (summarized in Figure \ref{fig:pipeline}).

The DeepCERES pipeline consists on the following steps :
\begin{itemize}
	\item Noise removal: The Spatially Adaptive Non-local means (SANLM) filter \cite{Manjon2010}  was used to reduce random noise naturally present in the images.
	\item Registration: Affine registration to MNI152 space (1 mm\textsuperscript{3} resolution). ANTs software \cite{Avants2009} was employed. 
	\item Inhomogeneity correction: The N4 bias correction method was used to correct the inhomogeneity of the images \cite{Tustison2010}.
	\item Intensity normalization: We normalized the T1 images by applying a piecewise linear tissue mapping based on the TMS method \cite{Manjon2008} as described in the study by \cite{Manjon2016}.
	\item Intracranial cavity volume (ICV) extraction: To compute normalized volumes, we segmented the ICV using the Deep ICE method \cite{Manjon2020}.
	\item Second inhomogeneity correction and intensity normalization: This was performed using the ICV extracted volume instead of the original image to further improve the image quality.
	\item Super-resolution: The T1 image was super-resolved to 0.125 mm\textsuperscript{3} resolution (factor 2) using an in-house ResNet \cite{He2015} like super-resolution network trained using the HCP dataset. This step generated a volume of 362x434x362 voxels with synthetic 0.125 mm\textsuperscript{3} resolution.
	\item Cropping: A crop of the cerebellar region of the super-resolved T1 image is made (crop size of 162x182x168 voxels). This region was estimated from the limits of the library's manual segmentations in the MNI152 space (including 10 voxel margin in each dimension to accommodate cerebellum location and size variability).
	\item Atlas generation: The case-specific atlas is calculated as previously described from the training library, using a network to estimate the deformation fields of the 20 most similar cases and a parallel implementation for the weighted label fusion.
	\item T2 Synthesis: We used a cropped version of a recently proposed full volume convolutional neural network for MR image synthesis \cite{Manjon2021}.
	\item Segmentation: Once the cropped T1, the synthetized T2 and the subject specific atlas are generated we run the proposed ensemble-based method to generate the segmentation.
	\item Report generation: Finally, to make the analysis of the results more user friendly we generate a pdf report (and a CSV file) summarising the volumetric results derived from the obtained segmentation. This report includes the volumes of the different lobes in absolute value and normalized to the ICV (asymmetric indexes are also supplied). If the user supplies the age and sex of the subject a comparison with the lifespan model is performed, highlighting if each specific structure is inside the population bounds for a given age and sex. 

\end{itemize}

We are aware that the outcomes for super-resolved T1 and synthetic T2 images may not match the quality of native multimodal high-resolution data. To estimate this performance drop, we applied the proposed pipeline to monomodal and down-sampled T1 data of the test set. The results are listed in Table \ref{tab:sinT1}. 

\begin{table}[H]
  \centering
    \begin{adjustbox}{width=\columnwidth,center}
 \begin{tabular}{C{0.33\linewidth}|C{0.33\linewidth}C{0.33\linewidth}}
\multicolumn{1}{C{0.33\linewidth}}{\textbf{Labels}} & \multicolumn{1}{C{0.33\linewidth}}{\textbf{Ensemble of models with original multimodal HR data }} & \multicolumn{1}{C{0.33\linewidth}}{\textbf{Ensemble of models with synthetic multimodal HR data}} \bigstrut[b]\\
\hline
\hline
whole cerebellum  & 0.9900 (0.0025 & 0.9877 (0.0022 \bigstrut\\
\hline
Right Lobule I-II & 0.8319 (0.0551 & 0.8308 (0.0574 \bigstrut[t]\\
Right Lobule III & 0.9132 (0.0218 & 0.9077 (0.0221 \\
Right Lobule IV & 0.9283 (0.0052 & 0.9217 (0.0086 \\
Right Lobule V & 0.9317 (0.0130 & 0.9278 (0.0141 \\
Right Lobule VI & 0.9558 (0.0079 & 0.9516 (0.0078 \\
Right Crus I & 0.9606 (0.0072 & 0.9574 (0.0054 \\
Right Crus II & 0.9516 (0.0122 & 0.9479 (0.0101 \\
Right Lobule VIIB & 0.9319 (0.0232 & 0.9282 (0.0235 \\
Right Lobule VIIIA & 0.9441 (0.0185 & 0.9404 (0.0190 \\
Right Lobule VIIIB & 0.9453 (0.0085 & 0.9425 (0.0085 \\
Right Lobule IX & 0.9443 (0.0263 & 0.9404 (0.0224 \\
Right Lobule X & 0.9452 (0.0083 & 0.9404 (0.0118 \\
Right Grey Matter & 0.9605 (0.0029 & 0.9567 (0.0030 \\
Left Lobules I-II & 0.8234 (0.0442 & 0.8253 (0.0685 \\
Left Lobule III & 0.9159 (0.0165 & 0.9121(0.0168 \\
Left Lobule IV & 0.9206 (0.0140 & 0.9161(0.0124 \\
Left Lobule V & 0.9240 (0.0136 & 0.9164(0.0161 \\
Left Lobule VI & 0.9556 (0.0065 & 0.9505 (0.0075 \\
Left Crus I & 0.9600 (0.0061 & 0.9547 (0.0086 \\
Left Crus II & 0.9544 (0.0053 & 0.9502 (0.0058 \\
Left Lobule VIIB & 0.9319 (0.0207 & 0.9309 (0.0192 \\
Left Lobule VIIIA & 0.9370 (0.0214 & 0.9328 (0.0210 \\
Left Lobule VIIIB & 0.9392 (0.0107 & 0.9356 (0.0108 \\
Left Lobule IX & 0.9459 (0.0136 & 0.9420 (0.0133 \\
Left Lobule X & 0.9436 (0.0121 & 0.9409 (0.0134 \\
Left White Matter & 0.9619 (0.0026 & 0.9577 (0.0026 \bigstrut[b]\\
\hline
Average    & 0.9330 (0.0386 & 0.9290(0.0392 \bigstrut[t]\\
\end{tabular}%
  
    \end{adjustbox}
    \vspace{0.1 cm}
    \caption{Impact of not having high resolution (HR) and multimodal data.}
  \label{tab:sinT1}%
\end{table}

It is evident that there is a reduction in performance (0.9330 vs 0.9290).  However, we believe that the results still remain competitive.

\begin{figure}[H]
    \centering
    \includegraphics[width=\textwidth]{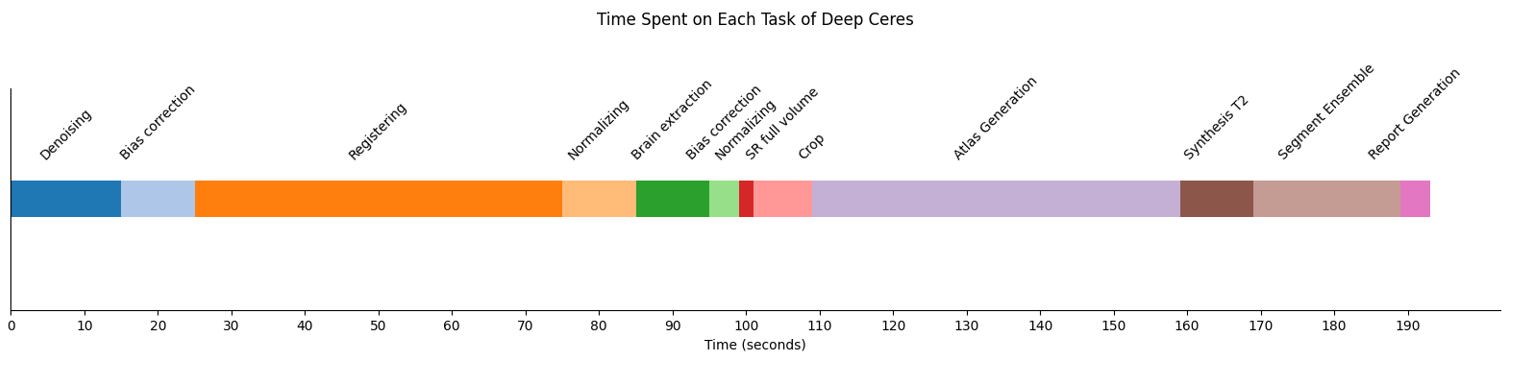}
    \caption{Temporal profile of DeepCERES method}
    \label{fig:temporal_profile}
\end{figure}

Finally, we assessed the efficiency of the proposed pipeline and its usability. We verified that the proposed pipeline can run for inference on a Titan Xp with 12 GB of memory. The complete pipeline has total temporal cost (from denoising to report generation) of approximately 3 minutes (details can be seen in Figure \ref{fig:temporal_profile}). The use of optimization-based registration tools (registration by ANTS to reach the MNI152 space) and the estimation and application of 20 deformation fields for atlas generation are the processes that make up most of the execution time of the method.

We will make DeepCERES available to the scientific community through our online system \href{https://volbrain.net/}{volBrain.net} .

\newpage
\section{Discussion}
In this paper we have proposed an ultra-high resolution multimodal cerebellum lobule segmentation method based deep learning. To train the proposed method we have created a semi-automatically segmented ultra-high resolution multimodal library. Thanks to the enhanced resolution of this training set a detailed delineation of the “arbor vitae” has been made possible allowing a more precise definition of the lobular grey matter. 

To be able to process the full cerebellar volume, the proposed method split the problem into two sub problems using a cascade approach. The validation of the cascade approach has been a crucial step in proving the effectiveness of our method in addressing a very complicated problem: segmentation in 27 structures with high-resolution multimodal volumes. 

We have also demonstrated that combining deep learning and classical machine learning approaches (multi-atlas) can be a successful strategy for enhancing performance and robustness in both architectures (using a priori knowledge through the atlas). Combining these two philosophies has made it possible to create more accurate models with a higher Dice index value, more efficient when compared to other more powerful models (with more parameters), and more robust with this a priori information.

The use of alternative models to the classic U-net network, such as the DPN model, has also been explored, achieving outstanding results despite having 5 times fewer parameters. Since it is commonly known that simpler models are less prone to overfit and, hence, generalise better, this information is crucial. However, the best outcomes in this experiment came from combining different models (U-net and DPN), demonstrating that the use of a bagging approach can help to reduce the classification error. 

Furthermore, the proposed approach has been subjected to several analyses in terms of robustness and generality, allowing us to improve its capabilities. After the study, a robust method is presented capable of generalizing over the lifespan of the human cerebellum. 

Despite the pseudo-quantitative nature of this comparison of this method with state-of-the-art alternatives due to the limitations of the data and the protocols employed, the average Dice indices are higher in the proposed method (0.870 vs 0.936), indicating a very competitive outcome.

Finally, we plan to make the proposed method openly accessible to the scientific community. To this end, a complete pipeline has been created to allow users to employ the proposed approach starting with a single standard-resolution T1 image.

\section{Conclusion}
In this study, we introduce a novel method for cerebellum lobule segmentation utilizing deep learning on ultra-high resolution multimodal MR images. Our experiments confirm the hypothesis that leveraging ultra-high resolution and multimodal data enhances the precision of cerebellum lobule segmentation. Furthermore, we demonstrate the efficacy of combining classical methods with deep learning, incorporating a specific atlas as a priori information.

We introduce a new pipeline, DeepCERES, ready to become publicly accessible through the online service \href{https://www.volbrain.net}{volBrain.net}. This pipeline is designed to seamlessly process standard-resolution T1 images, making it particularly valuable for analyzing numerous existing legacy datasets.

The upcoming availability of DeepCERES opens avenues for in-depth analysis of normal and pathological cerebellar patterns, shedding new light on this structure's involvement in various neurological diseases.

\newpage
\section*{Acknowledgements}
This work has been developed thanks to the project PID2020-118608RB-I00 (AEI/10.13039/501100011033) of the Ministerio de Ciencia e Innovacion de España. This work also benefited from the support of the project DeepvolBrain and HoliBrain of the French National Research Agency (ANR-18-CE45-0013 and ANR-23-CE45-0020-01). This study was achieved within the Laboratory of Excellence TRAIL ANR-10-LABX-57 for the BigDataBrain project. Moreover, we thank the Investments for the future Program IdEx Bordeaux (ANR-10- IDEX- 03- 02, HL-MRI Project), Cluster of excellence CPU and the CNRS.

Moreover, this work is based on multiple samples. We wish to thank all investigators of these projects who collected these datasets and made them freely accessible.
The C-MIND data used in the preparation of this article were obtained from the C-MIND Data Repository (accessed in February 2015) created by the C-MIND study of Normal Brain Development. This is a multisite, longitudinal study of typically developing children from ages newborn through young adulthood conducted by Cincinnati Children's Hospital Medical Center and UCLA A listing of the participating sites and a complete listing of the study investigators can be found at \href{https://research.cchmc.org/c-mind}{link}.

The NDAR data used in the preparation of this manuscript were obtained from the NIH-supported National Database for Autism Research (NDAR). NDAR is a collaborative informatics system created by the National Institutes of Health to provide a national resource to support and accelerate research in autism. The NDAR dataset includes data from the NIH Pediatric MRI Data Repository created by the NIH MRI Study of Normal Brain Development. This is a multisite, longitudinal study of typically developing children from ages newborn through young adulthood conducted by the Brain Development Cooperative Group A listing of the participating sites and a complete listing of the study investigators can be found at \href{http://pediatricmri.nih.gov/nihpd/info/participating_centers.html}{link}.

The ADNI data used in the preparation of this manuscript were obtained from the Alzheimer's Disease Neuroimaging Initiative (ADNI). The ADNI is funded by the National Institute on Aging and the National Institute of Biomedical Imaging and Bioengineering and through generous contributions from the following: Abbott, AstraZeneca AB, Bayer Schering Pharma AG, Bristol-Myers Squibb, Eisai Global Clinical Development, Elan Corporation, Genentech, GE Healthcare, GlaxoSmithKline, Innogenetics NV, Johnson \& Johnson, Eli Lilly and Co., Medpace, Inc., Merck and Co., Inc., Novartis AG, Pfizer Inc., F. Hoffmann-La Roche, Schering-Plough, Synarc Inc., as well as nonprofit partners, the Alzheimer's Association and Alzheimer's Drug Discovery Foundation, with participation from the U.S. Food and Drug Administration. Private sector contributions to the ADNI are facilitated by the Foundation for the National Institutes of Health (www.fnih.org). The grantee organization is the Northern California Institute for Research and Education, and the study was coordinated by the Alzheimer's Disease Cooperative Study at the University of California, San Diego. ADNI data are disseminated by the Laboratory for NeuroImaging at the University of California, Los Angeles.

The OASIS data used in the preparation of this manuscript were obtained from the OASIS project. See \href{http://www.oasis-brains.org/}{link} for more details. The AIBL data used in the preparation of this manuscript were obtained from the AIBL study of ageing. See \href{www.aibl.csiro.au} link for further details. The ICBM data used in the preparation of this manuscript. The IXI data used in the preparation of this manuscript were supported by the \href{http://www.brain-development.org/}{Brain Development}.

The ABIDE data used in the preparation of this manuscript were supported by ABIDE funding resources listed at \href{http://fcon\_1000.projects.nitrc.org/indi/abide/}{link}. ABIDE primary support for the work by Adriana Di Martino. Primary support for the work by Michael P. Milham and the INDI team was provided by gifts from Joseph P. Healy and the Stavros Niarchos Foundation to the Child Mind Institute, \href{http://fcon_1000.projects.nitrc.org/indi/abide/}{link}

\newpage
\bibliographystyle{unsrt}  
\bibliography{references}  
\newpage
\section*{Appendix}
\label{sec:Appendix}

\end{document}